\documentclass[a4paper]{article}
\usepackage[english]{babel}
\usepackage{amsmath,amsfonts,amssymb}
\usepackage{graphicx}

\begin{document}

\relax

\def\be{\begin{equation}}
\def\ee{\end{equation}}
\def\bs{\begin{subequations}}
\def\es{\end{subequations}}
\def\calm{{\cal M}}
\def\calk{{\cal K}}
\def\Hc{{\cal H}}
\def\lx{\lambda}
\def\sx{\sigma}
\def\ex{\epsilon}
\def\Lx{\Lambda}

\newcommand{\bett}{\tilde{\beta}}
\newcommand{\alpht}{\tilde{\alpha}}
\newcommand{\dv}{\delta{v}}
\newcommand{\dvvec}{\delta\vec{v}}
\newcommand{\phit}{\tilde{\phi}}
\newcommand{\Phit}{\tilde{\Phi}}
\newcommand{\Gt}{\tilde{G}}
\newcommand{\mt}{\tilde{m}}
\newcommand{\Mt}{\tilde{M}}
\newcommand{\nt}{\tilde{n}}
\newcommand{\Nt}{\tilde{N}}
\newcommand{\Bt}{\tilde{B}}
\newcommand{\Rt}{\tilde{R}}
\newcommand{\rt}{\tilde{r}}
\newcommand{\mut}{\tilde{\mu}}
\newcommand{\mub}{\bar{\mu}}
\newcommand{\vrm}{{\rm v}}
\newcommand{\tl}{\tilde t}
\newcommand{\ttt}{\tilde T}
\newcommand{\rhot}{\tilde \rho}
\newcommand{\ptt}{\tilde p}
\newcommand{\drho}{\delta \rho}
\newcommand{\dpp}{\delta p}
\newcommand{\dphi}{\delta \phi}
\newcommand{\drhot}{\delta {\tilde \rho}}
\newcommand{\dchi}{\delta \chi}
\newcommand{\A}{A}
\newcommand{\B}{B}
\newcommand{\mmu}{\mu}
\newcommand{\mnu}{\nu}
\newcommand{\ii}{i}
\newcommand{\jj}{j}
\newcommand{\jl}{[}
\newcommand{\jr}{]}
\newcommand{\ml}{\sharp}
\newcommand{\mr}{\sharp}

\newcommand{\da}{\dot{a}}
\newcommand{\db}{\dot{b}}
\newcommand{\dn}{\dot{n}}
\newcommand{\dda}{\ddot{a}}
\newcommand{\ddb}{\ddot{b}}
\newcommand{\ddn}{\ddot{n}}
\newcommand{\pa}{a^{\prime}}
\newcommand{\pn}{n^{\prime}}
\newcommand{\ppa}{a^{\prime \prime}}
\newcommand{\ppb}{b^{\prime \prime}}
\newcommand{\ppn}{n^{\prime \prime}}
\newcommand{\fda}{\frac{\da}{a}}
\newcommand{\fdb}{\frac{\db}{b}}
\newcommand{\fdn}{\frac{\dn}{n}}
\newcommand{\fdda}{\frac{\dda}{a}}
\newcommand{\fddb}{\frac{\ddb}{b}}
\newcommand{\fddn}{\frac{\ddn}{n}}
\newcommand{\fpa}{\frac{\pa}{a}}
\newcommand{\fpb}{\frac{\pb}{b}}
\newcommand{\fpn}{\frac{\pn}{n}}
\newcommand{\fppa}{\frac{\ppa}{a}}
\newcommand{\fppb}{\frac{\ppb}{b}}
\newcommand{\fppn}{\frac{\ppn}{n}}
\newcommand{\at}{\tilde{\alpha}}
\newcommand{\pt}{\tilde{p}}
\newcommand{\Ut}{\tilde{U}}
\newcommand{\phidot}{\dot{\phi}}
\newcommand{\rhb}{\bar{\rho}}
\newcommand{\pb}{\bar{p}}
\newcommand{\pbb}{\bar{\rm p}}
\newcommand{\kt}{\tilde{k}}
\newcommand{\kb}{\bar{k}}
\newcommand{\wt}{\tilde{w}}

\newcommand{\dA}{\dot{A_0}}
\newcommand{\dB}{\dot{B_0}}
\newcommand{\fdA}{\frac{\dA}{A_0}}
\newcommand{\fdB}{\frac{\dB}{B_0}}

\def\be{\begin{equation}}
\def\ee{\end{equation}}
\def\bs{\begin{subequations}}
\def\es{\end{subequations}}
\newcommand{\een}{\end{subequations}}
\newcommand{\ben}{\begin{subequations}}
\newcommand{\beq}{\begin{eqalignno}}
\newcommand{\eeq}{\end{eqalignno}}

\def \lta {\mathrel{\vcenter
     {\hbox{$<$}\nointerlineskip\hbox{$\sim$}}}}
\def \gta {\mathrel{\vcenter
     {\hbox{$>$}\nointerlineskip\hbox{$\sim$}}}}

\def\g{\gamma}
\def\mpl{M_{\rm Pl}}
\def\ms{M_{\rm s}}
\def\ls{l_{\rm s}}
\def\l{\lambda}
\def\m{\mu}
\def\n{\nu}
\def\a{\alpha}
\def\b{\beta}
\def\gs{g_{\rm s}}
\def\d{\partial}
\def\co{{\cal O}}
\def\sp{\;\;\;,\;\;\;}
\def\r{\rho}
\def\dr{\dot r}

\def\e{\epsilon}
\newcommand{\NPB}[3]{\emph{ Nucl.~Phys.} \textbf{B#1} (#2) #3}   
\newcommand{\PLB}[3]{\emph{ Phys.~Lett.} \textbf{B#1} (#2) #3}   
\newcommand{\ttbs}{\char'134}        
\newcommand\fverb{\setbox\pippobox=\hbox\bgroup\verb}
\newcommand\fverbdo{\egroup\medskip\noindent%
                        \fbox{\unhbox\pippobox}\ }
\newcommand\fverbit{\egroup\item[\fbox{\unhbox\pippobox}]}
\newbox\pippobox
\def\tr{\tilde\rho}
\def\lb{w}
\def\bbox{\nabla^2}
\def\mt{{\tilde m}}
\def\rct{{\tilde r}_c}

\def \lta {\mathrel{\vcenter
     {\hbox{$<$}\nointerlineskip\hbox{$\sim$}}}}
\def \gta {\mathrel{\vcenter
     {\hbox{$>$}\nointerlineskip\hbox{$\sim$}}}}

\noindent
\begin{flushright}

\end{flushright} 
\vspace{1cm}
\begin{center}
{ \Large \bf  Non-linear Matter Spectrum for a  Variable Equation of State \\}
\vspace{1cm}
{N. Brouzakis$^{(1)}$ and N. Tetradis$^{(2)}$}
\end{center}
\vspace{1.2cm}
(1) {\it Departament de F\'isica, Univeristat Aut\`onoma de Barcelona, 08193 Bellaterra, Barcelona, Spain} 
\\
(2) {\it Department of Physics, University of Athens, University Campus, Zographou 157 84, Greece}
\vspace{1cm}
\abstract{
We study the growth of matter perturbations beyond the linear level in
cosmologies in which the dark energy has a variable equation of state. 
The non-linear corrections
result in shifts in the positions of the maximum, minima and nodes of the spectrum within the range
of Baryon Acoustic Oscillations. These can be used in order to distinguish 
theories with different late-time variability of the equation of state. 
}

\newpage

\section{Introduction}

The evolution of the matter power spectrum as a function of redshift depends on the underlying 
cosmological scenario. Its form reflects the evolution of the Universe since
the time of matter-radiation equality. For given initial conditions, determined by the 
primordial spectrum (usually assumed to be scale invariant), 
the growth of matter perturbations can be used in order to
constrain the possible cosmological scenaria through the 
comparison of the resulting spectrum with the observed large-scale structure.
The most promiment feature of the matter spectrum is a series of peaks and valleys, characterized as 
Baryon Acoustic Oscillations (BAO). They originate in the period of recombination, and correspond to
sound waves in the relativistic plasma of that epoch. 

The chacteristic length scale of BAO is around
100 Mpc. The exact form of the matter power spectrum at such scales is not easy to compute precisely, because of 
the failure of linear perturbation theory to describe reliably the growth of the corresponding  
fluctuations under gravitational collapse.   
At length scales below about 10 Mpc, the evolution is highly
non-linear, so that only numerical N-body simulations can capture the dynamics of the formation 
of galaxies and clusters of galaxies. However, fluctuations with length scales of 
around 100 Mpc fall within the mildly non-linear regime, for which analytical methods
have been developed. We focus on scales in the range 50--200 Mpc, within which BAO are visible. 

The various analytical methods \cite{CrSc1}--\cite{matsubara} that have been developed 
in order to go beyond linear perturbation theory essentially amount to 
resummations of subsets of perturbative diagrams of arbitrarily high order, in a way analogous to the 
renormalization group (RG). (For a summary and comparison of the various methods, see \cite{carlson}.)
We follow the approach of  \cite{Max2}, named time-RG or TRG, which 
uses time as the flow parameter that describes the evolution of physical quantities,
such as the spectrum of perturbations. The method 
has been applied to ${\rm \Lambda CDM}$ and quintessence cosmologies \cite{Max2}, allowing for 
a possible coupling of dark energy to dark matter \cite{coupled}, as well as
models with massive neutrinos \cite{lesgourgues}. 

The purpose of the present work is to apply the formalism to the case of a variable equation of state of the
dark-energy component. The typical example of such a scenario links the dark energy to an evolving scalar field 
with non-zero potential \cite{scalar}. We use a simpler parametrization of the dark-energy sector, by assuming that its
energy-momentum tensor has the usual perfect-fluid form but with an equation of state that is a function of redshift:
$p/\rho=w(z)$. We assume the form of $w(z)$ employed in \cite{komatsu}: 
\be
w(a)=\frac{a \, \wt(a)}{a+a_{\rm trans}} -\frac{a_{\rm trans}}{a+a_{\rm trans}},
\label{param} \ee
where $a=1/(1+z)$ is the scale factor, and $\wt(a)=\wt_0+(1-a)\wt_a$. The value $a_{\rm trans}$ corresponds to
the ``transition epoch'', during which the dark energy ceases to have the form of a pure cosmological constant with
$w=-1$ and starts to evolve. At small redshifts eq. (\ref{param}) reproduces the commonly used linear form of $w(a)$
\cite{chevallier}
\be
w(a)=w_0+(1-a)w_a,
\label{linw} \ee
where $w_0$, $w_a$ can be expressed in terms of $\wt_0$, $\wt_a$.
Following \cite{komatsu}, we describe the various models through the present-day value of $w$, $w_0\equiv w(z=0)$, 
its first derivative, $w'=dw/dz|_{z=0}$, and $a_{\rm trans}$. For $a_{\rm trans} \ll 1$, we have 
$w_0\simeq \wt_0$ and $w'\simeq \wt_a$. A detailed discussion of this parametrization is given in appendix C
of \cite{komatsu}.

\begin{figure}[t]
\begin{minipage}{72mm}
\includegraphics[width=\linewidth,height=50mm]{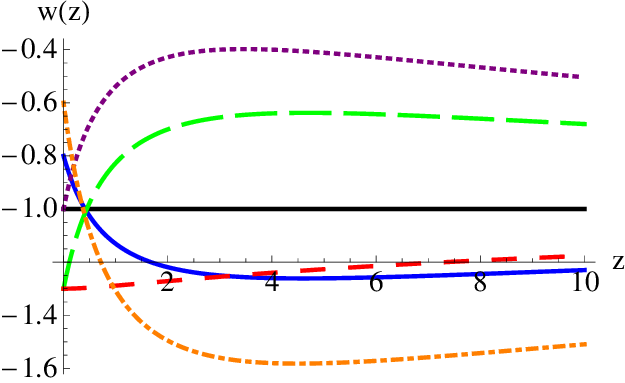}
\caption{The form of the equation of state $w(z)$ for ${\rm \Lambda CDM}$ and
a variety of models: 
$w_0=-1$, $w'=1$ (dotted), $w_0=-1.3$, $w'=1$ (long-dashed), $w_0=-1.3$, $w'=0$ (short-dashed), 
$w_0=-0.8$, $w'=-0.7$ (continuous), $w_0=-0.6$, $w'=-1.5$ (dash-dotted).}
\label{wz}
\end{minipage}
\hfil
\begin{minipage}{75mm}
\includegraphics[width=\linewidth,height=50mm]{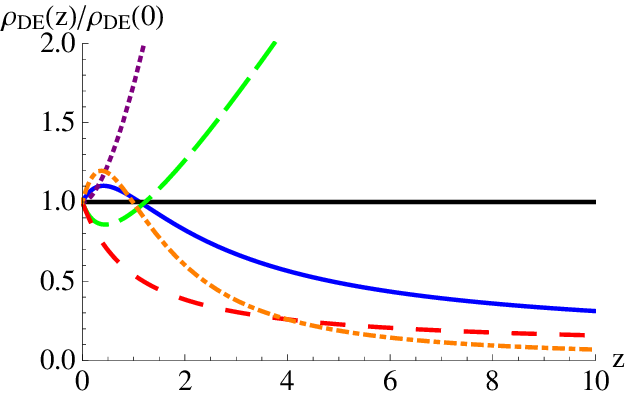}
\caption{The evolution of the dark-energy density for the models of fig. \ref{wz}.}
\label{dens}
\end{minipage}
\end{figure}

For our study we use values of $a_{\rm trans}$ such that the equation of state of the dark-energy sector is very close to
that of a cosmological constant during the time of recombination. In this way, we do not account for potentially interesting
models with significant amounts of evolving dark energy at early times. Instead, the emphasis of our study is on the
influence of the late-time evolution on the matter spectrum. Assuming that the dark-energy density is constant during recombination, 
with a negligible contribution to the total energy density, provides the most convenient framework in order to isolate
the late-time effects.

In fig. \ref{wz} we depict the form of $w(z)$ for several models, defined through the values of $w_0$ and $w'$. 
For all of them we assume $z_{\rm trans} = 10$. The equation of state varies significantly for $z<2$. For $z>2$ the
function $w(z)$ has a smooth evolution and approaches $w=-1$ asymptotically for large $z$. As a result, the early
cosmological evolution is similar to ${\rm \Lambda CDM}$ for all models. In fig. \ref{dens} we depict the evolution
of the dark-energy density as a function of redshift. We emphasize that the dark energy gives a significant contribution
to the total energy of the Universe only for redshifts $z\lta 2$, while at earlier times the dominant component is
dark matter. For example, at $z=10$ we have $\rho_{DE}/\rho_{DM}\lta 5\%$ for all models, while at $z=40$, where we 
set the initial conditions for the evolution of the power spectrum, we have $\rho_{DE}/\rho_{DM}\lta 0.5\%$.

In order to develop a self-consistent formalism beyond the linear level, we describe in the following section how the 
relevant evolution equations can be obtained, starting from the fundamental equations of motion and making the necessary
approximations. In our approach, the paramertrization of the equation of state by a variable $w(z)$ corresponds to
the presence of a quintessence field with an appropriate potential and a standard kinetic term. 
An immediate consequence of these assumptions is that the dark energy does not cluster at the sub-horizon level. 
Models in which substantial dark-energy clustering appears, as a consequence of either primordial isocurvature perturbations or 
non-trivial late-time dynamics \cite{gordonhu}, are beyond the scope of our study.

In the current analysis of data of galaxy surveys the non-linear corrections are accounted for in a rather crude fashion: 
The linear spectrum is multiplied by a Gaussian smoothing factor that 
results in the damping of small scales \cite{percival}. The dampling scale $D_{\rm damp}$ is set to $10\,h^{-1}$ Mpc.
Our purpose is to identify deformations of the spectrum beyond this approximation. The mode mixing generated by the
non-linear corrections affects the features of the spectrum (such as the extrema) in a non-uniform way. In this study we 
focus on the relative position of the maxima, minima and nodes of the spectrum in the BAO region. We determine
the shifts in their relative positions induced by the non-linear growth of perturbations. Such shifts can be used in order
to differentiate between models by lifting the degeneracies that persist in a cruder analysis. Our main aim is to determine
the magnitude of these effects. A careful comparison with actual data will require more sophisticated methods, such as
the Fisher matrix methodology \cite{fisher}.

\section{Formalism}

We concentrate on the evolution of fluctuations at subhorizon scales. 
We include contributions from all massive matter (baryonic and dark) in the total matter density $\rho(\tau,\vec{x})$.
The fundamental equations in the TRG approach are the 
continuity, Euler and Poisson equations. For an expanding background, they can be obtained
starting from the Einstein equations and the conservation of the energy-momentum tensor. 
This calculation has been performed in \cite{coupled} in the case that the dark energy arises through the
potential of an evolving scalar field $\phi$. We summarize the results, in order to 
deduce the relevant equations for the parametrization through a variable equation of state $w(z)$.
The details are given in the appendix A of ref. \cite{coupled}.

For the metric, we consider an ansatz of the form
\be
ds^2=a^2(\tau)\left[
\left(1+2\Phi(\tau,\vec{x}) \right)d\tau^2
-\left(1-2\Phi(\tau,\vec{x}) \right) d\vec{x}\, d\vec{x} \right].
\label{metric} \ee
We assume that the Newtonian potential $\Phi$ is weak, $\Phi \ll 1$, and  
that the field $\phi$ can be decomposed as
$\phi(\tau,\vec{x})=\bar{\phi}(\tau)+\dphi(\tau,\vec{x}),$
with $\dphi/\bar{\phi} \ll 1$. In general,
$\bar{\phi}={\cal O}(1)$ in units of the reduced Planck mass $M=(8\pi G)^{-1/2}$.
In the absence of a coupling between dark matter and dark energy, 
the magnitude of the fluctuations of $\phi$ is not 
expected to exceed that of the gravitational field $\Phi$.
Finally, the density can be decomposed as
$\rho(\tau,\vec{x})=\bar{\rho}(\tau)+\drho(\tau,\vec{x}).$
We allow for significant
density fluctuations, even though our analysis is not applicable when they
are much larger than the background density.  

For subhorizon perturbations with momenta $k\gg \Hc=\dot{a}/a$, 
the linear analysis predicts 
$|\dvvec| \sim (k/\Hc) \Phi \sim (\Hc/k)(\drho/\bar{\rho})$. 
A consistent expansion scheme can be obtained if we assume that 
$\Phi \ll |\dvvec | \ll 1$. 
Including the density perturbations, our assumptions can be summarized in the  
hierarchy of
scales: $\Phi \ll |\dvvec | \ll \drho/\bar{\rho}  \lesssim 1$. 
We assume that the fluctuations of the scalar field $\dphi/\bar{\phi}$ can be as large as 
the gravitational potential $\Phi$. However, we shall see in the following that, if there is no direct coupling between dark matter
and dark energy,  they are actually much smaller. 
As we are dealing
with subhorizon perturbations, it is consistent to make
the additional assumption that the spatial derivatives of $\Phi, \dphi$
dominate over their time derivatives. 
The predictions of the linear analysis allow us to make a more quantitative
statement. We assume that a spatial derivative acting on $\Phi$, $\dphi$ or 
$\dvvec$ increases the position of that quantity in the hierarchy by one 
level. In this sense $\vec{\nabla} \Phi$ is comparable to $\dvvec$, while
$\nabla^2 \Phi$ is comparable to $\bar{\rho}$. 

The equation of motion for the scalar field has been derived in ref. \cite{coupled}.
Assuming the hierarchy we described above, it can be split into an
equation for the homogeneous part:
\be
\ddot{\bar{\phi}}+2\frac{\dot{a}}{a}\dot{\bar{\phi}}+a^2\frac{dU(\bar{\phi})}{d\phi}=0,
\label{homphi} \ee
and one for the perturbation:
\be
\delta\ddot{\phi}+2\frac{\dot{a}}{a}\delta\dot{\phi}-\nabla^2\delta
\phi+a^2\frac{d^2U(\bar{\phi})}{d\phi^2}\dphi-4\dot{\bar{\phi}}\dot{\Phi}
+2 a^2\frac{dU(\bar{\phi})}{d\phi}\Phi
=0.
\label{pertphi} \ee
The terms with
time derivatives are subdominant according to our assumptions.
The scalar field $\bar{\phi}$ takes values of order $M$. 
For $\bar{\phi}$ to evolve at cosmological times, it must have a mass term
$d^2U(\bar{\phi})/d{\phi}^2 = {\cal O} (\Hc^2)$. We also have
$U(\bar{\phi}),\, \bar{\rho} =  {\cal O} (\Hc^2 M^2)$. It is natural to expect 
$dU(\bar{\phi})/d{\phi} = {\cal O} (\Hc^2 M)$. For subhorizon perturbations
with characteristic momenta $k\gg \Hc$, the inspection of eq. (\ref{pertphi}) demonstrates 
that $\delta\phi/\bar{\phi}$ is suppressed by $(\Hc/k)^2$
relative to $\Phi$. In the absence of a direct coupling between dark matter and the field $\phi$,
the field fluctuation $\delta\phi/\bar{\phi}$ is negligible within the hierarchy we have assumed.

We can conclude that the fluctuations in the dark-energy sector do not affect the
growth of matter fluctuations at scales much smaller than the horizon distance. 
If the dark energy does not couple directly to 
matter, it does not cluster sufficiently for its fluctuations to play a role in structure formation. 
This conclusion is consistent with the approximations underlying the essentially
Newtonian framework that we employ for the study of the spectrum. It must be pointed out 
that it is possible for significant fluctuations in the dark-energy sector to originate in the
primordial spectrum. Such a scenario is considered in ref. \cite{gordonhu}. As we have discussed in
the introduction, we focus on modifications of the matter power spectrum induced by the 
late-time cosmological evolution. For this reason, we do not discuss the possibility of initial
dark-energy perturbations and their subsequent role.
In summary, we consider models without significant dark-energy clustering. 
At the practical level, we assume $\delta\phi\simeq 0$ in the following, an approximation 
consistent with our essentially Newtonian treatment of perturbations. 
The quintessence field affects only the background evolution, in a way modelled by a
time-dependent equation of state parametrized by $w(z)$. It is straightforward to generalize our
approach for the case of significant dark-energy fluctuations, along the lines of ref. \cite{coupled}.

The equation of motion for the gravitational potential $\Phi$ can be obtained from
the first Einstein
equation. The leading terms give 
\be
\Hc^2=\left( \frac{\dot{a}}{a} \right)^2=
\frac{1}{3 M^2}\left[ a^2 \bar{\rho}+\frac{1}{2}\dot{\bar{\phi}}^2+a^2 U(\bar{\phi}) 
\right],
\label{hubble} \ee
while our assumptions reduce the equation for the perturbation to the Poisson equation for
the gravitational field $\Phi$:
\be
\nabla^2\Phi=\frac{1}{2M^2} a^2 \drho.
\label{gravpois} \ee

We now turn to equations derived from the conservation of the total
energy-momentum tensor $T^{\mu\nu}_{~~;\nu}=0$. For $\mu=0$, the
leading terms give
\be
\dot{\bar{\rho}}+3\Hc \bar{\rho}
=-\frac{\dot{\bar{\phi}}}{a^2}
\left( \ddot{\bar{\phi}}+2\Hc \dot{\bar{\phi}}+a^2\frac{dU(\bar{\phi})}{d\phi}
\right)=0,
\label{cons2} \ee
where we have employed eq. (\ref{homphi}).
The equation for the perturbation is more complicated. It can be simplified 
considerably through
our assumptions about the hierarchy of the various fields. We obtain
\be
\delta\dot{\rho}+3\Hc\drho+\vec{\nabla}\left[(\bar{\rho}+\drho)\dvvec
\right]= \frac{\dot{\bar{\phi}}}{a^2} \nabla^2 \dphi\simeq
0,
\label{cons3} \ee
where we have made use of our assumption $\delta\phi\simeq 0$.
For $\mu=i=1,2,3$ we obtain the generalization of the Euler equation for this
system. After eliminating higher-order terms 
and employing eqs. (\ref{homphi}),  (\ref{cons2}), (\ref{cons3}) we find
\be
\delta \dot{\vec{v}}
+\Hc\dvvec
+\left(\dvvec\, \vec{\nabla}\right)\dvvec
=-\vec{\nabla} \Phi.
\label{cons5} \ee

We have seen that,  if the dark energy has no direct coupling to matter, it influences the matter spectrum only through its effect on
the expansion rate. We can now rewrite the relevant equations using the variable equation of state $w(a)$.
We replace the equation of motion of $\bar{\phi}$ by the continuity equation for
the dark-energy density in the form
\be
\dot{\bar{\rho}}_{DE}+3(1+w(a))\,\Hc\, \bar{\rho}_{DE}=0.
\label{consde} \ee
For a specific quintessence model, the determination of the function $w(a)$ requires the integration of eq. (\ref{homphi}).
In our study, however, we shall assume that 
$w(a)$ is parametrized by eq. (\ref{param}). 
The Friedmann equation (\ref{hubble}) takes the form
\be
\Hc^2=\left( \frac{\dot{a}}{a} \right)^2=
\frac{1}{3 M^2}a^2\left(  \bar{\rho}+\bar{\rho}_{DE} 
\right).
\label{friedmann} \ee
In summary, the evolution of the background is determined by eqs. (\ref{cons2}), (\ref{consde}), (\ref{friedmann}), while
the growth of the matter fluctuations is described by eqs. (\ref{gravpois}), (\ref{cons3}), (\ref{cons5}).

The evolution equations are expressed in their most 
useful form in terms of the density contrast 
$\delta\equiv{\delta\rho}/{\bar{\rho}}\lesssim 1$
and 
$\theta(\textbf{k}, \tau)\equiv\vec{\nabla}\cdot\vec{\delta v}(\textbf{k}, \tau)$.
For the Fourier transformed quantities, 
we obtain 
\begin{equation}\label{delta}
\dot{\delta}(\textbf{k}, \tau)+\theta(\textbf{k}, \tau)
+\int d^{3}\textbf{k}_{1} \,d^{3}\textbf{k}_{2}\,
\delta_{D}(\textbf{k}-\textbf{k}_{1}-\textbf{k}_{2})\,
\alpht(\textbf{k}_{1}, \textbf{k}_{2})\,
\theta(\textbf{k}_{1}, \tau)\,
\delta(\textbf{k}_{2}, \tau)
=0,
\end{equation}
with 
$\alpht(\textbf{k}_{1}, \textbf{k}_{2})={\textbf{k}_{1}\cdot(\textbf{k}_{1}
+\textbf{k}_{2})}/{k_{1}^{2}},$
and
\begin{equation}\label{theta}
\dot{\theta}(\textbf{k}, \tau)+
\mathcal{H}\theta(\textbf{k}, \tau)
+\dfrac{3}{2}\mathcal{H}^{2}\Omega_M\delta
(\textbf{k}, \tau)
+\int d^{3}\textbf{k}_{1}\, d^{3}\textbf{k}_{2}\,
\delta_{D}(\textbf{k}-\textbf{k}_{1}-\textbf{k}_{2})\,
{\bett}(\textbf{k}_{1}, \textbf{k}_{2})\,
\theta(\textbf{k}_{1}, \tau)\,
\theta(\textbf{k}_{2}, \tau)=0,
\end{equation}
with 
$ \bett(\textbf{k}_{1}, \textbf{k}_{2})={(\textbf{k}_{1}+\textbf{k}_{2})^{2}
\textbf{k}_{1}\cdot\textbf{k}_{2}}/(2 k_{1}^{2}k_{2}^{2})$
and $\Omega_M\equiv\bar{\rho}a^2/(3\Hc^2)$.

We define the doublet
\begin{equation}
 \left(
\begin{array}{c}
\varphi_{1}(\textbf{k}, \eta)\\ \\ \varphi_{2}(\textbf{k}, \eta)
\end{array}
\right)
=e^{-\eta}\left(
\begin{array}{c}
\delta(\textbf{k}, \eta)\\ \\-\dfrac{\theta(\textbf{k}, \eta)}{\mathcal{H}}
\end{array}
\right),
\label{quadruplet}
\end{equation}
where $\eta=\ln a(\tau)$.  
This allows us to bring 
eqs. (\ref{delta}), (\ref{theta}) 
in the form \cite{CrSc1,Max1,Max2} 
\begin{equation}\label{arghh}
\partial_{\eta}\varphi_{a}(\textbf{k}, \eta)+\Omega_{ab}\varphi_{b}(\textbf{k}, \eta)
=e^{\eta}\gamma_{abc}(\textbf{k}, -\textbf{k}_{1}, -\textbf{k}_{2})
\varphi_{b}(\textbf{k}_{1}, \eta)\varphi_{c}(\textbf{k}_{2}, \eta).
\end{equation}
The indices $a,b,c$ take values $1,2$.
Repeated momenta are integrated over, while repeated indices are summed over. 
The functions $\gamma$, that determine effective vertices, 
are given in \cite{Max1,Max2}.
The non-zero components are
\begin{equation}
\begin{split}
\gamma_{121}(\textbf{k},\textbf{k}_{1}, \textbf{k}_{2})&
=\dfrac{\alpht(\textbf{k}_{1}, \textbf{k}_{2})}{2}
\delta_{D}(\textbf{k}+\textbf{k}_{1}+\textbf{k}_{2})
=\gamma_{112}(\textbf{k}, \textbf{k}_{2}, \textbf{k}_{1})\\
\gamma_{222}(\textbf{k},\textbf{k}_{1}, \textbf{k}_{2})
&=\bett(\textbf{k}_{1}, \textbf{k}_{2})\ 
\delta_{D}(\textbf{k}+\textbf{k}_{1}+\textbf{k}_{2}).
\end{split}
\label{vertex}
\end{equation}
The $\Omega$-matrix, which determines the linear evolution, is
\be
 \Omega(\eta)=\left(
\begin{array}{cc}
 1 & -1 
\\ \\
-\dfrac{3}{2}\Omega_M & 2+\dfrac{\mathcal{H}'}{\mathcal{H}} 
\end{array}
\right),
\label{omegafour}
\ee
where a prime denotes a derivative with respect to $\eta$.

The next step is to derive evolution equations for the power spectra.  
The spectrum and bispectrum are defined as
\begin{equation}
\begin{split}
 \langle\varphi_{a}(\textbf{k}, \eta)\varphi_{b}(\textbf{q}, \eta)\rangle
\equiv&
\delta_{D}(\textbf{k}+\textbf{q}) P_{ab}(\textbf{k}, \eta)\\
\langle\varphi_{a}(\textbf{k}, \eta)\varphi_{b}(\textbf{q}, \eta)\varphi_{c}(\textbf{p}, \eta)\rangle
\equiv&
\delta_{D}(\textbf{k}+\textbf{q}+\textbf{p}) B_{abc}(\textbf{k}, \textbf{q},\textbf{p},\eta)
\end{split}
\label{spectra} \end{equation}
Their evolution with time can be obtained through differentiation with respect to $\eta$ and use of eq. (\ref{arghh}). 
The essential approximation that we have to make in order to obtain a closed 
system is to neglect the effect of the trispectrum on the
evolution of the bispectrum. This gives \cite{Max1,Max2}
\begin{eqnarray}
\partial_{\eta}P_{ab}(\textbf{k}, \eta)&=&-\Omega_{ac}P_{cb}(\textbf{k}, 
\eta)-\Omega_{bc}P_{ac}(\textbf{k}, \eta)
\nonumber \\
&&+e^{\eta}\int d^{3}q \big[\gamma_{acd}(\textbf{k},-\textbf{q}, \textbf{q}-\textbf{k})
B_{bcd}(\textbf{k},-\textbf{q}, \textbf{q}-\textbf{k})
\nonumber \\
&&+\gamma_{bcd}(\textbf{k},
-\textbf{q}, \textbf{q}-\textbf{k})B_{acd}(\textbf{k},-\textbf{q}, \textbf{q}
-\textbf{k})\big],
\label{spectev1}\\
 \partial_{\eta}B_{abc}(\textbf{k},-\textbf{q}, \textbf{q}-\textbf{k})&=&
-\Omega_{ad}B_{dbc}(\textbf{k},-\textbf{q}, \textbf{q}-\textbf{k})-\Omega_{bd}B_{adc}
(\textbf{k},-\textbf{q}, \textbf{q}-\textbf{k})-\Omega_{cd}B_{abd}(\textbf{k},
-\textbf{q}, \textbf{q}-\textbf{k})
\nonumber \\
&&+2e^{\eta}\big[\gamma_{ade}(\textbf{k},-\textbf{q}, \textbf{q}-\textbf{k})
P_{db}(\textbf{q}, \eta)P_{ec}(\textbf{k}-\textbf{q}, \eta)
\nonumber \\
&&+\gamma_{bde}(-\textbf{q},\textbf{q}-\textbf{k}, \textbf{k})P_{dc}(\textbf{k}
-\textbf{q}, \eta)P_{ea}(\textbf{k}, \eta)
\nonumber \\
&&+\gamma_{cde}(\textbf{q}-\textbf{k}, 
\textbf{k}, -\textbf{q})P_{da}(\textbf{k}, \eta)P_{eb}(\textbf{q}, \eta)\big].
\label{spectev2}
\end{eqnarray}

The procedure of truncating the system of equations is commonly employed 
in the applications of the Wilsonian RG to field theory or statistical physics. (For a review, see \cite{erg}.)
The accuracy of the calculation can be determined either by enlarging the truncated system (by including the trispectrum,
for example) and examining the stability of the results, or by comparing with alternative methods. The second 
approach is often followed, because enlarging the truncation can increase the complexity of the calculation considerably.
In the case of the TRG, the agreement with results from N-body simulations for ${\rm \Lambda CDM}$ \cite{Max2}, 
quintessence and coupled quintessence \cite{coupled} is at the sub-percent level for the BAO range. 
We expect agreeement with alternative methods at the same level for the case of a variable equation of state 
we are studying. 
We shall see in the following that the features of the spectrum differ at the sub-percent level for various forms
of the equation of state. We emphasize, however, that in this work we make comparisons between results obtained always with our
method (e.g. for various forms of $w(z)$, at the linear or non-linear level), without reference to other approaches.  
As a result, the bigger systematic effects, the ones that generate the discrepancy between different methods, are 
expected to cancel out. 
In particular, 
for the parametrization of eq. (\ref{param}), we expect that the comparison we perform between models characterized by different 
values of $w_0$ and $w'$ can lead to the reliable identification of the relative variation of features in the spectrum, such
as the location of maxima, minima or nodes. 

\section{Results and discussion}

We consider models for which $w(z)$ is very close to $-1$ at high redshfit. We also 
assume a common normalization at high redshift for the matter spectra of all the models that we study.    
At the linear level, the location of the extrema of the spectrum at low redshift is expected to be independent of the form of $w(z)$.
The reason is that, at the linear level, the evolution of the matter spectrum amounts to multiplication by
an overall redshift-dependent factor (the linear growth rate). At the non-linear level, however, the shape of the
matter spectrum is modified through mode coupling during the growth of the fluctuations. 
The main purpose of our study is to determine the
magnitude of the shift of the peak location arising from the non-linear corrections for the various forms of $w(z)$.

\begin{figure}[t]
\begin{minipage}{72mm}
\includegraphics[width=\linewidth,height=50mm]{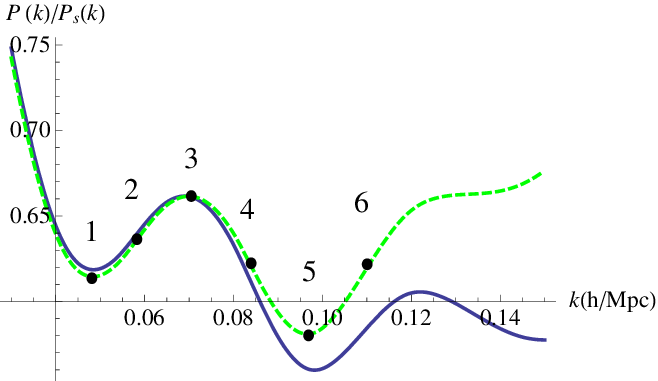}
\caption{Linear and non-linear spectra at $z=0$, for $w_0=-0.8$, $w'=-0.7$, normalized with respect to a smooth spectrum.}
\label{spectrum}
\end{minipage}
\hfil
\begin{minipage}{75mm}
\includegraphics[width=\linewidth,height=50mm]{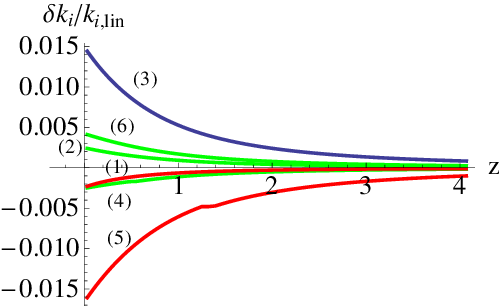}
\caption{The fractional shift of the maximum, minima and nodes of the non-linear spectrum, as a function of redshift, 
for $w_0=-0.8$, $w'=-0.7$.}
\label{shifts}
\end{minipage}
\end{figure}

The full system of eqs.~(\ref{spectev1}), (\ref{spectev2}) can be solved in a way analogous to 
that described in Appendix B of \cite{Max2}.  
We set the initial conditions for the integration of the evolution equations for the spectra at a redshift $z=40$. At such 
early times the evolution is linear to a very good approximation. Moreover, our assumptions about the form
of $w(z)$ imply that there is no appreciable amount of dark energy at such early times. As a result, the early 
evolution of the spectrum is identical to that in the ${\rm \Lambda}$CDM case.
We compute it by making use of the numerical code CAMB \cite{camb}. 
We assume that the primordial spectrum is scale invariant with spectral index $n=0.96$. 
We use the Cosmic Microwave Background (CMB) normalization for all the models we study: 
$\Delta_{\mathcal R}^2=2.46 \times 10^{-9}$ at $k_0=0.002/$Mpc.
We take the present-day Hubble parameter to be $H_0=70$ km/sec/Mpc,
the present-day total matter density $\Omega_M=0.27$ and the baryonic density
$\Omega_b=0.046$. Under the assumption that the dark energy is negligible before $z=40$, all the
models that we have studied have the same early Universe history. It is, therefore, consistent to assume
that they have the same power spectrum at $z=40$. The differences appear at low redshifts at which the
equation of state $w(z)$ deviates from a pure cosmological constant with $w=-1$. We take  
$a_{trans}=1/(1+z_{trans})$, with $z_{trans}=10$, for the ``transistion epoch", during which the dark energy 
starts deviating from a pure cosmological constant. The various models
predict different present-day spectra, age of the Universe, amplitude of matter fluctuations ($\sigma_8$), and
location of BAO peaks.

In fig. \ref{spectrum} we depict the linear (solid line) and non-linear (dashed line) matter power spectra at $z=0$ 
for a model with $w_0=-0.8$ and $w'=-0.7$. Following the commong practice, 
we have normalized both spectra with respect to a smooth spectrum. (The corresponding transfer function is
given by eq. (17) of ref. \cite{smooth}.) We have used a smooth function such that 
the non-linear spectrum is oscillatory around an almost constant value in the $k$-range of interest. On the same figure
we have indicated certain characteristic points of the spectrum: two minima (1 and 5), a maximum (3) and three nodes (2,4 and 6). 
In fig. \ref{shifts} we depict the evolution of these points as a function of redshift. At the linear level, no evolution is 
expected. For this reason, we plot the fractional difference in the location of these points from their constant 
values at the linear level. We observe that for large $z$ the deviation is negligible, because the spectrum is
linear to a good approximation. The deviations become significant below a redshift approximately equal to 3. 
We have checked that the choice of the smooth spectrum used for normalization 
does not influence appreciably the shift of the characteristic points.

\begin{figure}[t]
\begin{center}
\includegraphics[width=0.7\textwidth]{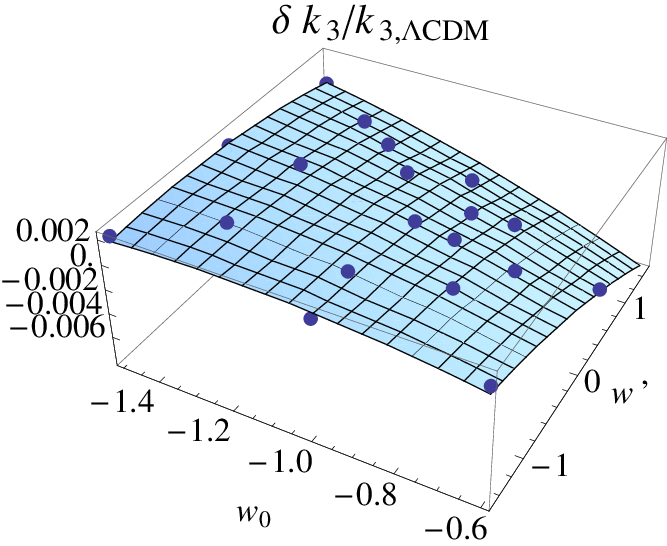}
\end{center}
\caption {The fractional shift of the first maximum from its location for ${\rm \Lambda}$CDM, as a function of
$w_0$ and $w'$, at a redshift $z=0.366$.} 
\label{threed}
\end{figure}

The change in the shape of the spectrum arising through the non-linear corrections can be used in order 
to distinguish different models. The ${\rm \Lambda}$CDM maximum
$k_{3,{\rm \Lambda CDM}}$ is shifted by $6.4\times 10^{-4}h$/Mpc when non-linear corrections are taken into account.
The shift is different for models with a variable equation of state. 
In fig. \ref{threed} we depict the location of the maximum of the non-linear spectrum as 
a function of $w_0$ and $w'$ at a redshift $z=0.366$. We plot the fractional deviation of the maximum
from its position in ${\rm \Lambda}$CDM (corresponding to the point $w_0=-1$, $w'=0$). 
We display several points that correspond to distinct
theories, as well as a smooth fit in order to guide the eye.
There are no points beyond the line starting at $(-0.6,0.5)$ and 
ending at $(-1.5,1.5)$. In that range of  the $(w_0,w')$ plane, and for our choice of 
$a_{trans}\simeq 0.091$, the values of the Hubble rate $H$ at $z=40$ differ by more than one per mille between the
various models. This implies that the effective equation of state $w(z)$ deviates from that of a pure
cosmological constant already at the initial time at which we have started the evolution of the spectrum. 
As we wish to disentangle such early-time effects from pure late-time ones, we do not investigate this part of the 
parameter range. 
It is apparent from fig. \ref{threed} that the late-time non-linear corrections to the spectrum generate differences at the 
per mille level for the location of the maximum in various dark-energy models. 

Similar behavior is observed for the location of the minima and nodes of the spectrum. 
The ratio of the position of the first minimum $k_1$ to that of the first maximum $k_3$,
as a function of $w_0$ and $w'$ at a redshift $z=0.366$, results in a surface 
with shape very similar to that in fig. \ref{threed}, but with opposite tilt.
The reason is that the 
first minimum is located in the part of the spectrum that is described accurately at the linear level, so that 
$k_1$ is essentially the same for all the theories we consider.  In table \ref{tab1} we give the values of $k_1/k_3$ for  
${\rm \Lambda}$CDM and four theories that correspond to the corner points in fig. \ref{threed}. 
The ratio of the position of the second minimum $k_5$ to $k_3$ displays enhanced variation, because
the non-linear corrections increase the value of $k_3$ but reduce the value of $k_5$. 
In table \ref{tab1} we give the values of $k_5/k_3$ for the same five theories. 

\begin{figure}[t]
\begin{minipage}{72mm}
\includegraphics[width=\linewidth,height=50mm]{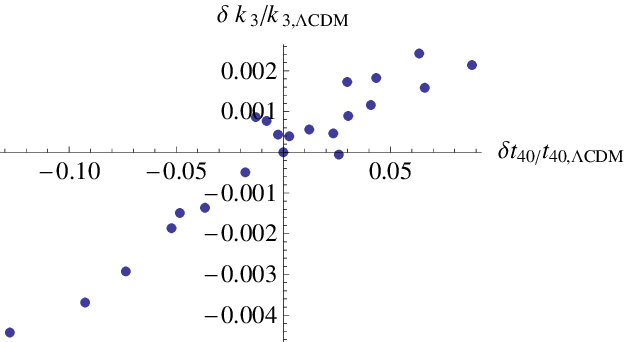}
\caption{The fractional shift of the first maximum from its location for ${\rm \Lambda}$CDM, as a function of
the age of the Universe.}
\label{shiftage}
\end{minipage}
\hfil
\begin{minipage}{75mm}
\includegraphics[width=\linewidth,height=50mm]{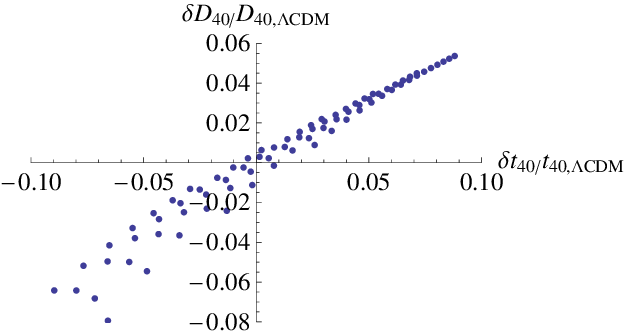}
\caption{Angular distance as a function of the age of the Universe for the class of models we consider.}
\label{lumtime}
\end{minipage}
\end{figure}

\begin{table}[b]
\caption{The location of mimima relative to the maximum for various theories.}
\label{tab1}
\begin{center}
\begin{tabular}{@{}llll@{}}
\hline
$w_0$ & $w'$ & $k_1/k_3$  & $k_5/k_3$\\
\hline
-1 & 0 & 0.6955 & 1.387 \\ \hline
-0.6 & -1.5 & 0.6948 & 1.394 \\ \hline
-1.5 & 1.5 & 0.6955 & 1.387 \\ \hline
-1.5 & -1.5 & 0.6939 & 1.377 \\ \hline
-0.6 & 0.5 & 0.6988 & 1.407 \\ \hline
\end{tabular}
\end{center}
\end{table}

There are certain degeneracies in the form of the non-linear spectrum, which are not immediately 
apparent in fig. \ref{threed}. In order to display them we depict in fig. \ref{shiftage} the same data points
as a function of the total time $t_{40}$ between $z=40$ and $z=0$, which is equal to the age of the Universe in a good
approximation.  We normalize this time with respect to its value for ${\rm \Lambda}$CDM. It is apparent that the 
shift of the maximum has an almost linear dependence on $t_{40}$. Models with distinct values of $w_0$ and $w'$, but
the same value of $t_{40}$, predict a similar shift of the maximum. This indicates that, if $\Omega_M$ is kept fixed in the
various models, as we 
have assumed throughout this work, the growth of non-linear perturbations depends mainly on the available time.  
An equivalent measure is the value of $\sigma_8$, as computed from the linear spectrum for each model. 
For fixed $\Omega_M$, the value of $\sigma_8$ increases with the available time. This results
in an almost linear dependence of the shift of the maximum on $\sigma_8$.

The various models that result in a shift  of the maximum very close to that for ${\rm \Lambda CDM}$ lie roughly on a line 
that starts at $(-0.6,-1.5)$ and ends at $(-1.5,1.3)$ on the $(w_0,w')$ plane. The same also happens for the shift of the 
second minimum, whose deviation from the value for ${\rm \Lambda CDM}$ can be as big as 1\%. 
The line between $(-0.6,-1.5)$ and $(-1.5,1.3)$  lies within the 
best-fit range obtained from a combination of the WMAP 5-year data, supernova data 
and the linear BAO specturm (fig. D1 of ref. \cite{komatsu}). The reason can be traced in the strong correlation between
the angular distance of a high-redshift source and the corresponding time.
In fig. \ref{lumtime} we depict the angular distance of a source at $z=40$ as a function 
of the total time $t_{40}$ between $z=40$ and $z=0$, for a large number of models with various values of
$w_0$ and $w'$. The same relation also holds for higher redshifts, as all the models have the same evolution before
$z=40$. 
It is apparent that for the class of models that we consider the time determines almost uniquely the
angular distance for high-redshift sources. As the time also determines the growth of the non-linear
effects in the matter spectrum, the appearance of the same degeneracy in the CMB and matter spectra 
is natural.

In conclusion, we have presented a calculation that demonstrates how the matter spectrum 
can be used in order to differentiate between cosmological models with a late-time deviation from ${\rm \Lambda}$CDM.
Features of the spectrum such as the location of extrema and nodes do not remain constant when non-linear 
corrections to the evolution are taken in to account. For example, non-linear corrections shift the location of
the first maximum towards larger values, while they decrease the value of the second minimum of the spectrum.

Our main quantitative result is that the relative differences in the features of the spectra for various models are at the 
sub-percent level. This makes them too small to be identified within the accuracy provided by the current data
of galaxy surveys.  However, effects such as the ones we discussed are calculable and 
can be used for comparisons as the precision will improve in the future. 
The most efficient method would make simultaneous use of the full spectrum, exploiting the fact
that the positions of extrema and nodes vary differently at the non-linear level. 
A simultaneous fit to the locations of maxima, minima and nodes at various redshifts
has the potential to provide a useful
method for differentiating between theories with late-time variation of the equation of state.

\section*{Acknowledgments}
We would like to thank M. Pietroni for useful discussions.
N.~B. and N.~T. are supported in part by the EU Marie Curie Network ``UniverseNet'' 
(MRTN--CT--2006--035863).  N.~T is also supported in part by the ITN network
``UNILHC'' (PITN-GA-2009-237920).




\begin{thebibliography}{999}

\bibitem{CrSc1}
  M.~Crocce and R.~Scoccimarro,
  Phys.\ Rev.\  D {\bf 73} (2006) 063519
  [arXiv:astro-ph/0509418].

 \bibitem{CrSc2}
  M.~Crocce and R.~Scoccimarro,
  Phys.\ Rev.\  D {\bf 73} (2006) 063520
  [arXiv:astro-ph/0509419];
  Phys.\ Rev.\  D {\bf 77} (2008) 023533
  [arXiv:0704.2783 [astro-ph]].

\bibitem{taruya}
  A.~Taruya and T.~Hiramatsu,
  arXiv:0708.1367 [astro-ph];
    T.~Hiramatsu and A.~Taruya,
  Phys.\ Rev.\  D {\bf 79} (2009) 103526
  [arXiv:0902.3772 [astro-ph.CO]];
    A.~Taruya, T.~Nishimichi, S.~Saito and T.~Hiramatsu,
  arXiv:0906.0507 [astro-ph.CO].


\bibitem{Max1}    
S.~Matarrese and M.~Pietroni,
  JCAP {\bf 0706} (2007) 026
  [arXiv:astro-ph/0703563].

\bibitem{Max2}
  M.~Pietroni,
  JCAP {\bf 0810} (2008) 036
  [arXiv:0806.0971 [astro-ph]].

\bibitem{coupled}
  F.~Saracco, M.~Pietroni, N.~Tetradis, V.~Pettorino and G.~Robbers,
  arXiv:0911.5396 [astro-ph.CO].

\bibitem{lesgourgues}
  J.~Lesgourgues, S.~Matarrese, M.~Pietroni and A.~Riotto,
  JCAP {\bf 0906} (2009) 017
  [arXiv:0901.4550 [astro-ph.CO]].

\bibitem{mcdonald}
  P.~McDonald,
  Phys.\ Rev.\  D {\bf 75} (2007) 043514
  [arXiv:astro-ph/0606028].

\bibitem{valeagas}
  P.~Valageas,
 Astron.\ Astrophys.\  {\bf 379} (2001) 8
  [arXiv:astro-ph/0107015];
 Astron.\ Astrophys.\  {\bf 382} (2002) 412
  [arXiv:astro-ph/0107126];
  Astron.\ Astrophys.\  {\bf 421} (2004) 23
  [arXiv:astro-ph/0307008];
 Astron.\ Astrophys.\  {\bf 465} (2007) 725
 [arXiv:astro-ph/0611849].

\bibitem{matsubara}
  T.~Matsubara,
  Phys.\ Rev.\  D {\bf 77} (2008) 063530
  [arXiv:0711.2521 [astro-ph]];
  Phys.\ Rev.\  D {\bf 78} (2008) 083519
  [Erratum-ibid.\  D {\bf 78} (2008) 109901]
  [arXiv:0807.1733 [astro-ph]].

\bibitem{carlson}
  J.~Carlson, M.~White and N.~Padmanabhan,
  Phys.\ Rev.\  D {\bf 80} (2009) 043531
  [arXiv:0905.0479 [astro-ph.CO]].

\bibitem{scalar}
 B.~Ratra and P.~J.~E.~Peebles,
  Phys.\ Rev.\  D {\bf 37} (1988) 3406;
  C.~Wetterich,
  Nucl.\ Phys.\  B {\bf 302} (1988) 668.

\bibitem{komatsu}
  E.~Komatsu {\it et al.}  [WMAP Collaboration],
  Astrophys.\ J.\ Suppl.\  {\bf 180} (2009) 330
  [arXiv:0803.0547 [astro-ph]].

\bibitem{chevallier}
  M.~Chevallier and D.~Polarski,
  Int.\ J.\ Mod.\ Phys.\  D {\bf 10} (2001) 213
  [arXiv:gr-qc/0009008];
  E.~V.~Linder,
  Phys.\ Rev.\ Lett.\  {\bf 90} (2003) 091301
  [arXiv:astro-ph/0208512].

\bibitem{gordonhu}
  C.~Gordon and W.~Hu,
  Phys.\ Rev.\  D {\bf 70} (2004) 083003
  [arXiv:astro-ph/0406496].

\bibitem{percival}
  W.~J.~Percival {\it et al.},
  Mon.\ Not.\ Roy.\ Astron.\ Soc.\  {\bf 401} (2010) 2148
  [arXiv:0907.1660 [astro-ph.CO]].

\bibitem{fisher}
  M.~Tegmark, A.~Taylor and A.~Heavens,
  Astrophys.\ J.\  {\bf 480} (1997) 22
  [arXiv:astro-ph/9603021];
 T.~D.~Kitching and A.~Amara,
  arXiv:0905.3383 [astro-ph.CO].



\bibitem{erg}
  J.~Berges, N.~Tetradis and C.~Wetterich,
  Phys.\ Rept.\  {\bf 363} (2002) 223
  [arXiv:hep-ph/0005122].

\bibitem{camb}
  A.~Lewis, A.~Challinor and A.~Lasenby,
  Astrophys.\ J.\  {\bf 538} (2000) 473
  [arXiv:astro-ph/9911177].

\bibitem{smooth}
  D.~J.~Eisenstein and W.~Hu,
  Astrophys.\ J.\  {\bf 496} (1998) 605
  [arXiv:astro-ph/9709112].








\end{thebibliography}
\end{document}